\documentclass[fleqn]{revtex4}

\usepackage{amsmath,graphicx}
\usepackage{verbatim,dcolumn}
\usepackage{multirow}
\newcommand\T{\rule{0pt}{3ex}} 

\begin{document}

\title{Higher-order corrections to the spin-orbit and spin-spin tensor interactions in HD$^+$}

\author{Mohammad Haidar$^1$\footnote[1]{Present affiliations: 1. Sorbonne Universit\'e, Laboratoire de Chimie Th\'eorique (LCT), 4 place Jussieu, F-75005 Paris, France 2. Sorbonne Universit\'e, CNRS, Universit\'e Paris Cit\'e, Laboratoire Jacques-Louis Lions (LJLL), 4 place Jussieu, F-75005 Paris, France 3. TotalEnergies, Tour Coupole, 2 Pl. Jean Millier, F-92078 Paris la D\'efense, France}, Vladimir I. Korobov$^{2}$, Laurent Hilico$^{1,3}$, and Jean-Philippe Karr$^{1,3}$}
\affiliation{$^1$Laboratoire Kastler Brossel, Sorbonne Universit\'e, CNRS, ENS-Universit\'e PSL, Coll\`ege de France, 4 place Jussieu, F-75005 Paris, France}
\affiliation{$^2$Bogoliubov Laboratory of Theoretical Physics, Joint Institute for Nuclear Research, Dubna 141980, Russia}
\affiliation{$^3$Universit\'e d'Evry-Val d'Essonne, Universit\'e Paris-Saclay, Boulevard Fran\c cois Mitterrand, F-91000 Evry, France}

\begin{abstract}
Improved values of hyperfine coefficients related to the electronic spin-orbit and electron-nucleus spin-spin tensor interactions in the HD$^+$ molecular ion are obtained through numerical calculation of relativistic corrections at the $m\alpha^6$ order and radiative corrections at the $m\alpha^7\ln(\alpha)$ order. The theoretical accuracy is improved by more than one order of magnitude. Some deviations with recent high-precision ro-vibrational spectroscopy experiments are observed, in contrast with the good agreement obtained in H$_2^+$.
\end{abstract}

\maketitle

\section{Introduction}

Laser spectroscopy of HD$^+$ has recently made a significant leap in precision, as a rotational transition~\cite{Alighanbari20} and two vibrational transitions~\cite{Patra20,Kortunov21} have been probed in the
Lamb-Dicke regime on ensembles of trapped and sympathetically cooled molecular ions. Transition frequencies were measured with relative uncertainties in the 10$^{-11}$-10$^{-12}$ range, and comparison with theoretical
predictions~\cite{Korobov17,Korobov21} was used to get improved determinations of the proton-electron mass ratio and constraints on a ``fifth force'' between the nuclei~\cite{Germann21}.

The hyperfine splitting of HD$^+$ rovibrational levels gives rise to 4, 10 or 12 sublevels for $L=0$, $L=1$ and $L \geq 2$ states respectively, where $L$ is the rotational quantum number~\cite{Bakalov06}. In the above-mentioned experiments, a few hyperfine components of the rovibrational transitions (between 2 and 6) were measured with sub-kHz uncertainties, from which a ``spin-averaged'' transition frequency was extracted using theoretical predictions of the hyperfine structure. A discrepancy between theory and experiment was evidenced in the hyperfine slitting of the $(v=0,L=3) \to (v'=9,L'=3)$ transition~\cite{Karr20}, which affects the uncertainty of the extracted spin-averaged transition frequency~\cite{Patra20,Koelemeij22}. Beyond that, constant progress in experimental accuracy opens the way to highly precise tests of the hyperfine structure theory. It is also worth noting that the HD$^+$ hyperfine structure is sensitive to the deuteron's electric quadrupole moment $Q_d$. A value of $Q_d$ was extracted from the rotational transition measurement~\cite{Alighanbari20}, although it is significantly less precise than that obtained from magnetic resonance measurements in $D_2$~\cite{Code71,Puchalski20}. For all these reasons, it is highly desirable to improve further the hyperfine structure theory in HD$^+$.

On the theoretical side, all the coefficients of the effective spin Hamiltonian introduced in~\cite{Bakalov06} have been calculated in the framework of the Breit-Pauli Hamiltonian with account of the electron's anomalous magnetic moment, which yielded a relative accuracy on the order of $\alpha^2$. We recently improved the precision of the largest hyperfine coefficients, i.e. the electron-proton and electron-deuteron spin-spin scalar interactions, to slightly below 1~ppm~\cite{Karr20}. The next largest coefficients are those related to the electron spin-orbit (denoted by $E_1$ in~\cite{Bakalov06}) and the electron-proton ($E_6$) and electron-deuteron ($E_7$) spin-spin tensor interactions. The effective Hamiltonian describing $m\alpha^6$-order relativistic corrections to these interactions was derived in~\cite{Korobov20} in the framework of nonrelativistic quantum electrodynamics (NRQED), and preliminary numerical results were given. In~\cite{Haidar22}, the theory was further refined by including radiative corrections at the following order ($m\alpha^7\ln(\alpha)$). The spin-orbit and spin-spin interaction coefficients were calculated numerically in the H$_2^+$ ion for a range of rovibrational states, and good agreement with available rf spectroscopy data was observed. In the present work, we report on the numerical calculation of these corrections in HD$^+$, which improves the theoretical accuracy of the $E_1$, $E_6$ and $E_7$ coefficients by almost one order of magnitude with respect to Ref.~\cite{Bakalov06}. We then compare our theoretical predictions of the hyperfine splitting with available experimental data~\cite{Alighanbari20,Patra20,Kortunov21}. Significant deviations are evidenced, possible causes of which are discussed in the last section.

\section{Numerical results} \label{sec-results}

Higher-order corrections to the spin-orbit and spin-spin tensor interaction coefficients in the hydrogen molecular ions were studied in our previous works~\cite{Korobov20,Haidar22}. The expressions of relativistic corrections at orders $m\alpha^6$ can be found in~\cite{Haidar22}, Eqs.~(17)-(24), and radiative corrections at the  $m\alpha^7\ln(\alpha)$ order are given in Eqs.~(27) and (29-35). In Tables~\ref{E1table-hdplus}~, \ref{E6results}, and \ref{E7results}, we report the results of numerical calculations of these correction terms for a few rovibrational states of HD$^+$ of direct interest for experiments~\cite{Alighanbari20,Patra20,Kortunov21,Zhong15}. Complete results for the rovibrational states $(1 \leq L \leq 4, 0 \leq v \leq 9)$ are given in the Appendix~\ref{SM}. These numerical calculations are performed following the methods presented in~\cite{Korobov20,Haidar22} relying on a variational expansion of the three-body wavefunction in terms of exponentials of interparticle distances, with pseudo-randomly chosen exponents~\cite{Thakkar77,Korobov00}.

The numerical and theoretical uncertainties were discussed in~\cite{Haidar22} in the context of the H$_2^+$ ion, and the main conclusions remain valid for the results presented here. Briefly, the numerical uncertainty of $E_1$ (resp. $E_6, E_7$) is estimated to be smaller than 10~Hz (resp. 1~Hz), see~\cite{Haidar22} from a study of convergence of a few terms. The overall uncertainty is dominated by yet unevaluated nonlogarithmic $m\alpha^7$-order contributions, and is estimated to about one third of the total contribution of order $m\alpha^7\ln(\alpha)$. The resulting uncertainties amount to 3-4~ppm for the electronic spin-orbit interaction and 2~ppm for tensor interactions respectively. This represents an improvement by factors of about~15 for $E_1$, and~25 for $E_6$ and $E_7$, over the results of Ref.~\cite{Bakalov06}.

\begin{table} [h!]
\small
\begin{tabular}{|@{\hspace{1mm}}c@{\hspace{1mm}}|@{\hspace{1mm}}c@{\hspace{1mm}}|@{\hspace{1mm}}c@{\hspace{1mm}}|@{\hspace{1mm}}c@{\hspace{1mm}}|@{\hspace{1mm}}c@{\hspace{1mm}}
|@{\hspace{1mm}}c@{\hspace{1mm}}|@{\hspace{1mm}}c@{\hspace{1mm}}|@{\hspace{1mm}}c@{\hspace{1mm}}|@{\hspace{1mm}}c@{\hspace{1mm}}
|@{\hspace{1mm}}c@{\hspace{1mm}}|@{\hspace{1mm}}c@{\hspace{1mm}}|@{\hspace{1mm}}c@{\hspace{1mm}}|@{\hspace{1mm}}c@{\hspace{1mm}}|@{\hspace{1mm}}c@{\hspace{1mm}}|}
\hline
$(L,v)$ &$\displaystyle  E_1^{(BP)}$& $\displaystyle\mathcal{U}_{Y_1}$ & $\displaystyle\mathcal{U}_{W}$ & $\displaystyle\mathcal{U}_{CM}$ & $\displaystyle\Delta E_{so\hbox{-}H_B}$ &$\displaystyle\Delta E_{so\hbox{-}so}^{(1)}$& $\displaystyle\Delta E_{so\hbox{-}ret}$ & $\displaystyle
\Delta  E_1^{(6)}$ & $\displaystyle\mathcal{U}_{Y_1}$ & $\displaystyle\mathcal{U}_{q^2}$ & $\displaystyle \Delta E_{so\hbox{-}H_{(5\ln)}}$ & $\displaystyle
\Delta  E_1^{(7\ln)}$ & $\displaystyle E_1$ (this work) \\[2mm]
\hline
(1,0) & 31984.645 & 1.170 & -2.736 & 0.021 & 2.088 & 0.312 & 0.257 & 1.112 & -0.026 & 0.045 & -0.367 & -0.348 & 31985.41(12) \\
(1,1) & 30280.027 & 1.110 & -2.631 & 0.027 & 1.986 & 0.299 & 0.243 & 1.035 & -0.025 & 0.044 & -0.345 & -0.326 & 30280.74(11) \\
(1,5) & 24076.867 & 0.893 & -2.205 & 0.042 & 1.592 & 0.233 & 0.193 & 0.747 & -0.020 & 0.036 & -0.268 & -0.252 & 24077.36(8)  \\
(1,6) & 22643.449 & 0.834 & -2.097 & 0.044 & 1.498 & 0.219 & 0.181 & 0.679 & -0.019 & 0.035 & -0.251 & -0.235 & 22643.89(8)  \\
(3,0) & 31627.353 & 1.156 & -2.694 & 0.019 & 2.042 & 0.308 & 0.255 & 1.086 & -0.026 & 0.045 & -0.360 & -0.341 & 31628.10(11) \\
(3,9) & 18270.560 & 0.680 & -1.732 & 0.043 & 1.158 & 0.182 & 0.146 & 0.478 & -0.015 & 0.028 & -0.198 & -0.185 & 18270.85(6) \\
\hline
\end{tabular}
\caption{Corrections to the spin-orbit interaction coefficient $E_1$ for a few rovibrational states of HD$^+$ (in kHz). The leading-order (Breit-Pauli) value $E_1^{(BP)}$ (Ref.~\cite{Bakalov06}) is given in column 2. Columns 3-5 and 6-8 are respectively the first-order and second-order contributions [Eqs.~(18) and (19) of~\cite{Haidar22}] at the  m$\alpha^6$ order, and the total correction at this order,  $\Delta  E_1^{(6)}$, is given in column 9. Columns 10-11 are the first-order contributions at the $m\alpha^7\ln(\alpha)$ order [Eqs.~(27) and (29) of~\cite{Haidar22}] whereas column 12 is the second-order term [Eq.~(32)] of~\cite{Haidar22}]. The total correction at this order, $\Delta  E_1^{(7\ln)}$, is given in column 13. The last column is our value of $E_1$. Its estimated uncertainty (equal to one third of $\Delta  E_1^{(7\ln)}$) is indicated between parentheses.\label{E1table-hdplus}}
\end{table}

\begin{table} [h!]
\small
\begin{tabular}{|@{\hspace{1mm}}c@{\hspace{1mm}}|@{\hspace{1mm}}c@{\hspace{1mm}}|@{\hspace{1mm}}c@{\hspace{1mm}}|@{\hspace{1mm}}c@{\hspace{1mm}}
|@{\hspace{1mm}}c@{\hspace{1mm}}|@{\hspace{1mm}}c@{\hspace{1mm}}|@{\hspace{1mm}}c@{\hspace{1mm}}|@{\hspace{1mm}}c@{\hspace{1mm}}|}
\hline
$(L,v)$ & $\displaystyle E_6^{(BP)}$ & $\displaystyle\mathcal{U}_W^{(2)}$&$\displaystyle\mathcal{U}_{CM}^{(2)}$ & $\displaystyle\Delta E_{ss\hbox{-}H_B}^{(2)}$ & $\displaystyle\Delta E_6^{(6)}$ &  $\displaystyle\Delta E_6^{(7\ln)}$ & $\displaystyle E_6$ (this work) \\[2mm]
\hline
(1,0) & 8611.112 & -0.806  & 0.093  & 0.955  & 0.242  & -0.055  & 8611.299(18) \\
(1,1) & 8136.686 & -0.770  & 0.090  & 0.905  & 0.225  & -0.052  & 8136.859(17) \\
(1,5) & 6421.232 & -0.632  & 0.079  & 0.726  & 0.172  & -0.041  & 6421.364(14) \\
(1,6) & 6027.810 & -0.599  & 0.076  & 0.678  & 0.154  & -0.039  & 6027.925(13) \\
(3,0) & 948.5222 & -0.0883 & 0.0101 & 0.1042 & 0.0260 & -0.0060 & 948.5421(20) \\
(3,9) & 538.9906 & -0.0544 & 0.0070 & 0.0593 & 0.0119 & -0.0035 & 539.9991(12) \\
\hline
\end{tabular}
\caption{Corrections to the electron-proton spin-spin tensor interaction coefficient $E_6$ for a few rovibrational states of HD$^+$ (in kHz). The leading-order (Breit-Pauli) value $E_6^{(BP)}$ (Ref.~\cite{Bakalov06}) is given in column 2. Columns 3-4 and 5 are respectively the first-order and second-order contributions [Eqs.~(22) and (23) of~\cite{Haidar22}] at the $m\alpha^6$ order. The total correction at this order,  $\Delta  E_6^{(6)}$, is given in column 6. Column 7 is the second-order contribution at the m$\alpha^7\ln(\alpha)$ order [Eq.~(35) of~\cite{Haidar22}]. The last column is our value for $E_6$. Its estimated uncertainty (equal to one third of $\Delta  E_6^{(7\ln)}$) is indicated between parentheses. \label{E6results} }
\end{table}

\begin{table} [ht!]
\small
\begin{tabular}{|@{\hspace{1mm}}c@{\hspace{1mm}}|@{\hspace{1mm}}c@{\hspace{1mm}}|@{\hspace{1mm}}c@{\hspace{1mm}}|@{\hspace{1mm}}c@{\hspace{1mm}}
|@{\hspace{1mm}}c@{\hspace{1mm}}|@{\hspace{1mm}}c@{\hspace{1mm}}|@{\hspace{1mm}}c@{\hspace{1mm}}|@{\hspace{1mm}}c@{\hspace{1mm}}|}
\hline
$(L,v)$ & $\displaystyle E_7^{(BP)}$ & $\displaystyle\mathcal{U}_W^{(2)}$&$\displaystyle\mathcal{U}_{CM}^{(2)}$ & $\displaystyle\Delta E_{ss\hbox{-}H_B}^{(2)}$ & $\displaystyle\Delta E_7^{(6)}$ &  $\displaystyle\Delta E_7^{(7\ln)}$ & $\displaystyle E_7$ (this work) \\[2mm]
\hline
(1,0) & 1321.7673 & -0.1237 & 0.0142 & 0.1467 & 0.0371 & -0.0085 & 1321.7960(28) \\
(1,1) & 1248.9359 & -0.1182 & 0.0138 & 0.1390 & 0.0346 & -0.0080 & 1248.9624(27) \\
(1,5) &  985.5868 & -0.0971 & 0.0121 & 0.1115 & 0.0264 & -0.0063 &  985.6069(21) \\
(1,6) &  925.1895 & -0.0919 & 0.0115 & 0.1040 & 0.0237 & -0.0060 &  925.2072(20) \\
(3,0) &  145.5938 & -0.0136 & 0.0016 & 0.0160 & 0.0040 & -0.0009 &  145.5969(3)  \\
(3,9) &   82.7237 & -0.0083 & 0.0010 & 0.0091 & 0.0018 & -0.0005 &   82.7250(2)  \\
\hline
\end{tabular}
\caption{Same as Table~\ref{E6results}, for the electron-deuteron spin-spin tensor coefficient $E_7$. \label{E7results}}
\end{table}

\section{Comparison with experiments and discussion} \label{comparison}

\subsection{Theoretical hyperfine splitting of rovibrational transitions} \label{theor-hfs}

These results allow us to improve theoretical predictions of the hyperfine splitting of rovibrational transitions that were recently measured with high accuracy. These are obtained by diagonalizing the effective spin Hamiltonian of Ref.~\cite{Bakalov06}, which comprises nine coefficients $E_1 \ldots E_9$. The values of $E_1$, $E_6$ and $E_7$ are taken from the present work. For the largest coefficients, i.e the electron-proton and electron-deuteron spin-spin scalar interaction coefficients (``Fermi'' contact interaction), $E_4$ and $E_5$, we use the very precise values computed in~\cite{Karr20}. Finally, the smaller coefficients $E_2$, $E_3$, $E_8$ and $E_9$ are calculated in the framework of the Breit-Pauli Hamiltonian~\cite{Bakalov06}. We have updated the $E_9$ coefficient, which corresponds to the effect of the deuteron's quadrupole moment $Q_d$, with the latest and most precise determination of $Q_d$~\cite{Puchalski20}. The values of all the coefficients used here can be found in the Appendix~\ref{hfs-coeff}. The fractional uncertainties of $E_4$ and $E_5$ are estimated to 0.93 and 0.57~ppm, respectively~\cite{Karr20}, and those of $E_2$, $E_3$ and $E_8$ are taken as equal to~$\alpha^2 \simeq 53$~ppm. Finally, for $E_9$ we add quadratically the $\alpha^2$ theoretical uncertainty and the 82~ppm uncertainty of $Q_d$~\cite{Puchalski20}, yielding a total fractional uncertainty of 98~ppm.

The theoretical uncertainties of hyperfine intervals $f_{ij} = f_j - f_i$, where $i$ ($j$) denotes the lower (upper) hyperfine states, and $f_i$, $f_j$ their respective hyperfine shifts, are then calculated by propagating the uncertainties of the hyperfine coefficients, using the derivatives
\begin{equation} \label{def-sensit}
\gamma_{i,k} = \frac{\partial f_{i}}{\partial E_k} \, , 1 \leq k \leq 9.
\end{equation}
The values of these derivatives for the hyperfine states involved in the $v=0 \to 9$ transition~\cite{Patra20} are given in the Appendix~\ref{sensit}. For the states relevant to the rotational transition and to the $v=0 \to 1$ transition, they can be found in~\cite{Alighanbari20} and the Supplementary information of~\cite{Kortunov21}, respectively.

It is important to take correlations into account in order to get reliable uncertainty estimates. There are strong correlations between theoretical errors of a given coefficient $E_k$ in different rovibrational states, which can be understood as follows. First, these errors are dominated by yet uncalculated QED contributions (numerical uncertainties being negligibly small), which are given by expectation values of the same effective operators, evaluated with different wavefunctions. Second, the uncalculated terms are corrections to the bound electron and essentially depend on the electronic part of the wavefunction, which varies only slowly with the rovibrational state. Correlations may also exist between errors of different coefficients $E_k$, $E_l$ for the same rovibrational state if they originate from similar QED terms. The hypotheses we have adopted for our uncertainty estimates are presented in the Appendix~\ref{correl}.

\begin{table} [h!]
\small
\begin{tabular}{|@{\hspace{1mm}}c@{\hspace{1mm}}|@{\hspace{1mm}}c@{\hspace{1mm}}|@{\hspace{1mm}}c@{\hspace{1mm}}|@{\hspace{1mm}}c@{\hspace{1mm}}|@{\hspace{1mm}}r@{\hspace{1mm}}|@{\hspace{1mm}}r@{\hspace{1mm}}
|@{\hspace{1mm}}r@{\hspace{1mm}}|@{\hspace{1mm}}r@{\hspace{1mm}}|@{\hspace{1mm}}c@{\hspace{1mm}}|}
\multicolumn{9}{c}{(a) $(v=0,L=0) \to (v'=1,L'=1)$} \\
\hline
$i$ & $FSJ \rightarrow F'S'J'$ & $j$ & $FSJ \rightarrow F'S'J'$ & $f_{ij}^{exp}$~\cite{Kortunov21} & $f_{ij}^{theor}$~\cite{Kortunov21} & $f_{ij}^{theor}$ (this work) & $\Delta_{ij}$ & $\Delta_{ij}/\sigma_c$ \\
\hline
12 & $122 \rightarrow 121$ & 16 & $122 \rightarrow 123$ & 41 294.06(32) & 41 293.81(44) & 41 293.66(12) & 0.40 & 1.2 \\
\hline
\multicolumn{9}{c}{(b) $(v=0,L=3) \to (v'=9,L'=3)$} \T \\
\hline
$i$ & $FSJ \rightarrow F'S'J'$ & $j$ & $FSJ \rightarrow F'S'J'$ & $f_{ij}^{exp}$~\cite{Patra20} & $f_{ij}^{theor}$~\cite{Karr20} & $f_{ij}^{theor}$ (this work) & $\Delta_{ij}$ & $\Delta_{ij}/\sigma_c$ \\
\hline
$F=0$ & $014 \rightarrow 014$ & $F=1$ & $125 \rightarrow 125$ & 178 254.4(9) & 178 246.2(1.8) & 178 245.89(28) & 8.5 & 9.0 \\
\hline
\end{tabular}
\begin{tabular}{|@{\hspace{1mm}}c@{\hspace{1mm}}|@{\hspace{1mm}}c@{\hspace{1mm}}|@{\hspace{1mm}}c@{\hspace{1mm}}|@{\hspace{1mm}}c@{\hspace{1mm}}|@{\hspace{1mm}}c@{\hspace{1mm}}|@{\hspace{1mm}}c@{\hspace{1mm}}|}
\multicolumn{6}{c}{(c) $(v=0,L=0) \to (v'=0,L'=1)$} \T \\
\hline
coefficient & $E_k^{exp}$ & $E_k^{theor}$ & $E_k^{theor}$ (this work) & $\Delta E_k$ & $\Delta E_k/\sigma_c$  \\
\hline
$E_1$ & 31 984.9(1) & 31 985.76(35) \cite{Korobov20} & 31 985.41(12)  & -0.5  & -3.3 \\
$E_6$ & 8 611.17(5) & 8 611.1(5) \cite{Bakalov06}    & 8 611.299(18)  & -0.13 & -2.4 \\
$E_7$ & 1 321.72(4) & 1 321.77(7) \cite{Bakalov06}   & 1 321.7960(28) & -0.08 & -2.0 \\
\hline
\end{tabular}
\caption{Comparison between experimental and theoretical hyperfine intervals (unit: kHz). In parts (a) and (b), hyperfine components are identified by their label (following the notations of the original publications~\cite{Patra20,Kortunov21}) and by the hyperfine states in the lower and upper rovibrational states (columns 1-4). Experimental values and previous theoretical predictions of the hyperfine intervals are given in column 5 and 6. Our predictions are shown in column 7. The deviations $\Delta_{ij} = f_{ij}^{exp} - \Delta_{ij}^{theor}$ are given in kHz in column 8, and in units of the combined uncertainty $\sigma_c = ( [ u(f_{ij}^{exp})]^2 + [ u(f_{ij}^{theor}) ]^2)^{1/2}$ in the last column. In part (c), column 2 contains the values of the $E_1$, $E_6$ and $E_7$ hyperfine coefficients extracted from the experimental data of Ref.~\cite{Alighanbari20} using a least-squares adjustment described in Appendix~\ref{LSA}. Previous theoretical predictions are given in column 3. Our predictions are shown in column 4. Deviations $\Delta E_k = E_k^{exp} - E_k^{theor}$ are given in kHz in column 5, and in units of the combined uncertainty in the last column.\label{comparison-exp}}
\end{table}

\subsection{Comparison with experimental data}

A comparison between experiment and theory is shown in Table~\ref{comparison-exp}. Regarding the hyperfine splitting of the $(v=0,L=0) \to (v'=1,L'=1)$ transition~\cite{Kortunov21} and of the $(v=0,L'=3) \to (v=9,L'=3)$ transition~\cite{Patra20}, the theoretical precision is significantly improved, and previous conclusions remain essentially valid: reasonable agreement between theory and experiment is observed in the first case, whereas a clear discrepancy appears in the second case, becoming even more significant (9 combined standard deviations) due to the reduced theoretical error bar. The cause of this discrepancy is currently unknown, but it is clear that it has no relationship with the corrections calculated in the present work, which shift the theoretical prediction by only -0.3~kHz with respect to the previous evaluation~\cite{Karr20}.

We now discuss the rotational transition~\cite{Alighanbari20}, which represents the most stringent test of our improved values of the $E_1$, $E_6$, and $E_7$ coefficients, in view of the higher absolute precision of the measurements and of their much lower sensitivity to the Fermi coefficients $E_4$ and $E_5$ (due to strong cancellation between the lower and upper states). Comparison between theory and experiment is more involved in this case, as six hyperfine components of this transition have been measured, from which fifteen hyperfine intervals can be deduced (the values of which are, of course, partially redundant). A detailed comparison for these fifteen intervals is shown in the Appendix~\ref{detail-rot}, revealing significant discrepancies for several lines. These data are however difficult to interpret, and a useful overview of the results can be obtained by extracting experimental values of $E_1$, $E_6$, and $E_7$ from a least-squares adjustment of the experimental data, whose results are shown in the last line of Table~\ref{comparison-exp}. The difference between experimental and theoretical values is above 3~$\sigma_c$ for the spin-orbit coefficient, and 2~$\sigma_c$ for spin-spin tensor coefficients. Details of the least-squares adjustment procedure can be found in the Appendix~\ref{LSA}.

\subsection{Discussion}

The observed deviations could be due to a problem in the theory, or in the experiment, or both; let us discuss here the first of these possibilities.

The largest discrepancy in terms of absolute magnitude (8.5~kHz, or a fractional difference of $4.8\!\times\!10^{-5}$) is in the hyperfine splitting of the $v = 0 \to 9$ transition. As discussed in~\cite{Karr20} it points towards the largest hyperfine coefficients $E_4$ and/or $E_5$, because the error that  would be required in the other coefficients to explain the discrepancy is much larger than the expected order of magnitude of any corrections beyond the Breit-Pauli Hamiltonian. However, a mistake in the calculation of the electron-proton spin-spin coefficient $E_4$ would also affect at a similar level the corresponding coefficient (denoted by $b_F$) in H$_2^+$, where agreement with experiments at a sub-kHz level is observed~\cite{Karr20}. This suggests that the problem might come from the electron-deuteron interaction coefficient $E_5$ (note that this hypothesis is not contradictory with the reasonably good agreement obtained for the $v = 0 \to 1$ transition frequency, because the latter only weakly depends on $E_5$).

What could then be missing in the theory of the electron-deuteron spin-spin interaction ? The difference between proton and deuteron cases resides in the nuclear structure and recoil corrections; these have been determined phenomenologically from the ground-state hyperfine splitting of the hydrogen and deuterium atoms~\cite{Karr20} under the assumptions that they are state-independent, i.e., they can be described by a contact (delta-function) term. Possible errors induced by this approximation are linked to the existence of state-dependent corrections. The largest such term is a recoil correction of order $(Z\alpha)^2(m/M)E_F$~\cite{Bodwin88}, which contributes to the ground-state hyperfine splitting of the deuterium atom at a level of only 39~Hz. An error on the electron-deuteron spin-spin interaction at a level of $\sim 10$~kHz thus seems extremely unlikely.

Regarding the rotational transition, the deviations between theory and experiment could be explained by errors in the spin-orbit ($E_1$) and spin-spin tensor coefficients ($E_6$, $E_7$) coefficients, as revealed by the adjustment described in the previous section (see last line of Table~\ref{comparison-exp}). However, any error in the calculation of $E_1$ for the $(L=1,v=0)$ would affect the $(L=1,v=1)$ level at a similar level. If one assumes, for example, that $E_1$ is shifted by -0.5~kHz for both states with respect to our present theoretical value, this would improve the agreement for the rotational transition, but increase the deviation from 0.4~kHz to about 0.9~kHz for the $v = 0 \to 1$ transition. Moreover, it would affect the value of the corresponding coefficient (denoted by $c_e$) in H$_2^+$, which we calculated in our previous work~\cite{Haidar22}, and cause significant tension with rf spectroscopy data. Overall, it appears that available measurements are not fully consistent with each other.

Apart from errors or missed contributions in theoretical hyperfine coefficients, another possible cause of deviations between theory and experiment is the incompleteness of the effective spin Hamiltonian introduced in Ref.~\cite{Bakalov06}. Only the terms that appear at the leading order (Breit-Pauli Hamiltonian) have been included so far, but other spin couplings appear at the order $m\alpha^6$ and higher. The largest one is the spin-spin contact interaction between proton and deuteron ($\mathbf{I}_p\!\cdot\!\mathbf{I}_d$). This term is already present in the Breit-Pauli Hamiltonian, but was not included in the effective spin Hamiltonian because the associated coupling coefficient is proportional to the proton-deuteron delta function expectation value, which is negligibly small due to the strong Coulomb repulsion. However, a larger coupling appears at the $m\alpha^6$ order due to the second-order contribution mediated by the electron. This contribution was studied in~\cite{Korobov22}, where it was shown that the coupling constant is on the order of 100~Hz, but shifts the hyperfine components of the rotational transition by less than 2~Hz due to cancellation between the lower and upper levels.

In conclusion, we did not identify any contribution that could potentially have the required order of magnitude to explain the observed discrepancies, and there is a strong need for additional measurements of the HD$^+$ hyperfine structure to give new insight on this problem.

\section*{Acknowledgements} The authors thank J. C. J. Koelemeij for helpful discussions.

\appendix

\section{Other coefficients of the effective spin Hamiltonian} \label{hfs-coeff}

We give in Table~\ref{table-hfs-coeff} the values of all the coefficients of the HD$^+$ effective spin Hamiltonian (see Eq.~(3) of~\cite{Bakalov06}) in the rovibrational states considered in Sec.~\ref{comparison} (see Table~\ref{comparison-exp}).

\begin{table}[h!]
\begin{tabular}{ |p{1cm}|p{1.5cm}|p{1cm}|p{2.5cm}|p{2.3cm}|p{1cm}|p{1cm}|}
 \hline
$(v,L)$ & $E_2$   & $E_3$  & $E_4$       & $E_5$       & $E_8$  & $E_9$ \\
\hline
$(0,0)$ &            &        & 925 394.159(860) & 142 287.556(84) &        &       \\
$(0,1)$ & -31.345(2) & -4.809 & 924 567.718(859) & 142 160.670(84) & -3.057 & 5.660 \\
$(1,1)$ & -30.463(2) & -4.664 & 903 366.501(839) & 138 910.266(82) & -2.945 & 5.653 \\
$(0,3)$ & -30.832(2) & -4.733 & 920 479.981(855) & 141 533.075(83) & -0.335 & 0.612 \\
$(9,3)$ & -21.304(1) & -3.225 & 775 706.122(721) & 119 431.933(73) & -0.219 & 0.501 \\
\hline
\end{tabular}
\caption{Hyperfine coefficients for a few rovibrational states of HD$^+$ (in kHz). The values of $E_4$, $E_5$ (resp.~$E_2$, $E_3$, $E_8$) are taken from~\cite{Karr20} (resp.~\cite{Bakalov06}). The value of $E_9$ has been updated with the latest determination of the deuteron's quadrupole moment~\cite{Puchalski20}. Uncertainties are discussed in Sec.~\ref{theor-hfs} of the main text. \label{table-hfs-coeff}}
\end{table}

\section{Dependence of hyperfine shifts on the hyperfine coefficients} \label{sensit}
The values of the derivatives $\gamma_{i,k}$ (as defined in Eq.~(\ref{def-sensit})), for the hyperfine states probed in spectroscopy of the $(L=3,v=0)\to(L=3,v=9)$ transition~\cite{Patra20}, are given in Table~\ref{table-sensit}. These quantities are required to estimate the uncertainty of the theoretical hyperfine interval.

\begin{table}[h!]
\begin{tabular}{ |p{2.25cm}|p{1.25cm}|p{1cm}|p{1cm}|p{1.25cm}|p{1.25cm}|p{1cm}|p{1.25cm}|p{1.25cm}|p{1.25cm}|}
 \hline
$(L,v,F,S,J)$ & $\gamma_{1,v}$  & $\gamma_{2,v}$&$\gamma_{3,v}$&$\gamma_{4,v}$&$\gamma_{5,v}$& $\gamma_{6,v}$&$\gamma_{7,v}$&$\gamma_{8,v}$&$\gamma_{9,v}$\\
\hline
(3,0,0,1,4) & -0.414 & 0.424 & 2.986 & -0.726 & -0.215 &  0.274 &  3.079 & -3.328 & -7.400 \\
(3,9,0,1,4) & -0.359 & 0.372 & 2.984 & -0.730 & -0.201 &  0.193 &  2.474 & -2.722 & -7.385 \\
(3,0,1,2,5) &  1.50  & 1.50  & 3.00  &  0.250 &  0.500 & -7.50  & -15.00 & -15.00 & -7.50  \\
(3,9,1,2,5) &  1.50  & 1.50  & 3.00  &  0.250 &  0.500 & -7.50  & -15.00 & -15.00 & -7.50  \\
\hline
\end{tabular}
\caption{Sensitivity coefficients for the hyperfine states involved in the $v=0-9$ transition in HD$^+$. Here $\gamma_{n,v} = \partial E_{hfs}(v,L,F,S,J)/\partial E_n$ . Note that for the $F=1$ level, the sensitivities are the same in $v=0$ and $v=9$, which is due to $(F=1,S=2,J=5)$ being a ``pure'' state. \label{table-sensit}}
\end{table}

\section{Correlations between the hyperfine coefficients} \label{correl}

In this Appendix, we present and justify our hypotheses regarding correlations between theoretical errors in the hyperfine coefficients, which are summarized in Table~\ref{table-correl}.

Let us first discuss the correlations between errors on a given coefficient $E_k$ in different rovibrational states. In the case of the largest hyperfine coefficients $E_4$ and $E_5$, related to the Fermi contact interaction between the electron and both nuclei, uncertainties and correlations can be controlled very well, thanks to the fact that these interactions have been studied in great depth in the context of the hydrogen and deuterium ground-state hyperfine splitting. This has allowed us to estimate the unevaluated QED terms from their value in the hydrogen (deuterium) atom ground state using the LCAO approximation. The uncertainties of $E_4$ and $E_5$ are conservatively taken as equal to the full estimated term~\cite{Karr20}. Then, for the evaluation of uncertainties of hyperfine shifts in rovibrational transitions, we assume perfect correlations between different rovibrational states, i.e., the uncertainty of a transition frequency is obtained by subtracting the uncertainties due to the upper and lower states. We have tested this assumption by applying it to lower-order terms that have been actually calculated by us, and found that the uncertainty estimated in this way matches well the magnitude of the actual correction.

For other hyperfine coefficients, we have used cruder uncertainty estimates based on the expected order of magnitude of the largest unevaluated QED terms. For example, for coefficients calculated at the Breit-Pauli, we estimate the relative uncertainty to be $\alpha^2$. In this case, it is harder to correctly evaluate the degree of correlation between theory errors, but all correlations are expected to be positive. In this work, we have assumed that the theory errors of the lower and upper levels are uncorrelated, which provides an upper limit of uncertainties on transition frequencies.

Let us now discuss correlations between errors in different coefficients $E_k$, $E_l$ for the same rovibrational state. Such correlations occur if the uncalculated QED terms that limit the theoretical precision are of the same nature. This is the case for the $E_2$ and $E_3$ coefficients that correspond to the proton ($E_2$) and deuteron ($E_3$) spin-orbit interactions. Both coefficients have been calculated in the framework of the Breit-Pauli Hamiltonien~\cite{Bakalov06}, and their theoretical errors are associated with $(Z\alpha^2)$-order relativistic corrections. These corrections will be described by the same effective potentials, the only difference being the substitution between proton and deuteron. Similarly, errors in the $E_6$ and $E_7$ coefficients that correspond to the electron-proton ($E_6$) and electron-deuteron ($E_7$) spin-spin tensor interactions are caused by the same QED term, that is the nonlogarithmic $m\alpha^7$-order radiative corrections (see Sec.~\ref{sec-results}). We thus assume perfect correlations between theory errors of $E_2$ and $E_3$, as well as between $E_6$ and $E_7$.

The case of the electron-proton ($E_4$) and electron-deuteron ($E_5$) Fermi contact interactions is slightly more involved. Indeed, errors in these coefficients originate from both nonrecoil QED corrections, which will have the same expressions apart from substitution of proton and deuteron, and from recoil corrections, which depend on the nucleus~\cite{Karr20}. We assume that errors associated with nonrecoil (resp. recoil) terms are fully correlated (resp. uncorrelated), which yields a correlation coefficient.
\begin{equation}
r(E_4,E_5) = \frac{u_{non-rec}(E_4) u_{non-rec}(E_5)} {u_{tot}(E_4) u_{tot}(E_5)} = 0.4016.
\end{equation}
Here, $u_{non-rec}(E_k)$ is the uncertainty of $E_k$ due to nonrecoil QED corrections, and $u_{tot}(E_k)$ its total uncertainty. These uncertainties are estimated to $u_{non-rec}(E_k) = 0.47 \times 10^{-6} E_k^{(F)}$, $u_{tot}(E_4) = 0.93 \times 10^{-6} E_4^{(F)}$, and $u_{tot}(E_5) = 0.59 \times 10^{-6} E_5^{(F)}$, where $E_k^{(F)}$ is the value of $E_k$ at the leading-order ($m\alpha^4$), see~\cite{Karr20} for details.

\begin{table} [h!]
\small
\begin{tabular}{|@{\hspace{2mm}}c@{\hspace{2mm}}|@{\hspace{2mm}}c@{\hspace{2mm}}|@{\hspace{2mm}}c@{\hspace{2mm}}|}
\hline
First coefficient & Second coefficient & Correlation coefficient \\
\hline
$E_k (v,L)$ & $E_k (v',L')$ & $1$ if $k = 4$ or $5$; $0$ otherwise. \\
\hline
$E_k (v,L)$ & $E_l (v,L)$ & \begin{tabular}{@{}c@{}} $1$ if $(k,l)=(2,3)$, $(3,2)$, $(6,7)$, or $(7,6)$;\\ $0.4016$ if $(k,l) = (4,5)$ or $(5,4)$; $0$ otherwise. \end{tabular} \\
\hline
$E_k (v,L)$ & $E_l (v',L')$ & \begin{tabular}{@{}c@{}} $0.4016$ if $(k,l)=(4,5)$ or $(5,4)$; $0$ otherwise. \end{tabular} \\
\hline
\end{tabular}
\caption{Summary of our assumptions on correlations between errors in the theoretical hyperfine coefficients. \label{table-correl}}
\end{table}

\section{Detailed comparison between theory and experiment for the hyperfine splitting of the rotational transition} \label{detail-rot}

A full comparison between the experimental results of Ref.~\cite{Alighanbari20} and our theoretical predictions is shown in Table~\ref{full-comparison-exp}.

\begin{table} [h!]
\small
\begin{tabular}{|@{\hspace{1mm}}c@{\hspace{1mm}}|@{\hspace{1mm}}c@{\hspace{1mm}}|@{\hspace{1mm}}c@{\hspace{1mm}}|@{\hspace{1mm}}c@{\hspace{1mm}}|@{\hspace{1mm}}r@{\hspace{1mm}}|@{\hspace{1mm}}r@{\hspace{1mm}}
|@{\hspace{1mm}}r@{\hspace{1mm}}|@{\hspace{1mm}}r@{\hspace{1mm}}|@{\hspace{1mm}}c@{\hspace{1mm}}|}
\multicolumn{9}{c}{$(v=0,L=0) \to (v'=0,L'=1)$} \\
\hline
$i$ & $FSJ \rightarrow F'S'J'$ & $j$ & $FSJ \rightarrow F'S'J'$ & $f_{ij}^{exp}$~\cite{Alighanbari20} & $f_{ij}^{theor}$~\cite{Alighanbari20} & $f_{ij}^{theor}$ (this work) & $\Delta_{ij}$ & $\Delta_{ij}/\sigma_c$ \\
\hline
12 & $122 \rightarrow 121$ & 14 & $100 \rightarrow 101$ & 24 134.211(75) & 24 134.5(1.1) & 24 134.465(23) & -0.254 & -3.2 \\
12 & $122 \rightarrow 121$ & 16 & $011 \rightarrow 012$ & 31 073.752(43) & 31 073.7(1.2) & 31 074.102(55) & -0.350 & -5.0 \\
12 & $122 \rightarrow 121$ & 19 & $122 \rightarrow 123$ & 43 283.419(54) & 43 283.4(2.2) & 43 284.10(12)  & -0.677 & -5.0 \\
12 & $122 \rightarrow 121$ & 20 & $122 \rightarrow 122$ & 44 944.338(72) & 44 944.6(1.9) & 44 945.289(64) & -0.951 & -9.9 \\
12 & $122 \rightarrow 121$ & 21 & $111 \rightarrow 112$ & 44 996.486(61) & 44 996.4(1.9) & 44 997.14(11)  & -0.652 & -5.4 \\
14 & $100 \rightarrow 101$ & 16 & $011 \rightarrow 012$ &  6 939.541(66) &  6 939.2(1.2) &  6 939.636(42) & -0.095 & -1.2 \\
14 & $100 \rightarrow 101$ & 19 & $122 \rightarrow 123$ & 19 149.208(74) & 19 148.8(2.0) & 19 149.63(11)  & -0.423 & -3.2 \\
14 & $100 \rightarrow 101$ & 20 & $122 \rightarrow 122$ & 20 810.127(88) & 20 810.1(1.8) & 20 810.823(62) & -0.696 & -6.5 \\
14 & $100 \rightarrow 101$ & 21 & $111 \rightarrow 112$ & 20 862.275(79) & 20 861.9(2.0) & 20 862.673(91) & -0.398 & -3.3 \\
16 & $011 \rightarrow 012$ & 19 & $122 \rightarrow 123$ & 12 209.667(41) & 12 209.7(1.5) & 12 209.994(72) & -0.327 & -4.0 \\
16 & $011 \rightarrow 012$ & 20 & $122 \rightarrow 122$ & 13 870.586(62) & 13 870.9(1.4) & 13 871.187(43) & -0.601 & -8.0 \\
16 & $011 \rightarrow 012$ & 21 & $111 \rightarrow 112$ & 13 922.734(49) & 13 922.7(1.0) & 13 923.037(51) & -0.303 & -4.3 \\
19 & $122 \rightarrow 123$ & 20 & $122 \rightarrow 122$ &  1 660.919(70) &  1 661.2(2.6) &  1 661.193(99) & -0.274 & -2.2 \\
19 & $122 \rightarrow 123$ & 21 & $111 \rightarrow 112$ &  1 713.067(59) &  1 713.0(0.7) &  1 713.042(25) &  0.025 &  0.4 \\
20 & $122 \rightarrow 122$ & 21 & $111 \rightarrow 112$ &     52.148(76) &     51.8(1.9) &     51.850(75) &  0.298 &  2.8 \\
\hline
\end{tabular}
\caption{Comparison between experimental and theoretical hyperfine intervals (unit: kHz). All definitions are identical to those of Table~\ref{comparison-exp}. \label{full-comparison-exp}}
\end{table}

\section{Least-squares adjustment of the spin-orbit and spin-spin tensor coefficients in the $(v=0,L=1)$ state from experimental data} \label{LSA}

The frequencies of the hyperfine components of the rotational transition (measured in~\cite{Alighanbari20}) can be written as a sum of the ``spin-averaged'' transition frequency, $f_{SA}$ and a hyperfine shift. In~\cite{Alighanbari20} (see also~\cite{Koelemeij22}), a value of $f_{SA}$ is extracted from the experimental data. Here, our goal is instead to extract experimental values of three hyperfine coefficients, $E_1$, $E_6$, and $E_7$. We do this in a way that is completely independent of the value of $f_{SA}$, which we describe in the following.

From the six measured transition frequencies, we deduce five independent hyperfine intervals by choosing one of the lines, $f_i$, as reference and computing the differences $f_{ij} = f_j - f_i$ for $j \neq i$. We then perform a least-square adjustment taking into account experimental and theoretical uncertainties. The latter are due to uncertainties in hyperfine coefficients (other than $E_1$, $E_6$, $E_7$), and are accounted for by additive corrections $\delta(E_k(v,L))$ that are treated as additional adjusted parameters, as done in the CODATA adjustments~\cite{Mohr00}. Correlations between theoretical errors are taken as described in the Appendix~\ref{correl}. Experimental uncertainties are assumed to be uncorrelated. For illustration, the input data used in the adjustment are shown in Table~\ref{table-LSA} for the case where the line labeled 12 is chosen as reference line.

\begin{table} [h!]
\small
\begin{tabular}{@{\hspace{1mm}}l@{\hspace{1mm}}}
\hline
$f_{14} - f_{12} = 24\,134.211(75)$~kHz \\
$f_{16} - f_{12} = 31\,073.752(43)$~kHz \\
$f_{19} - f_{12} = 43\,283.419(54)$~kHz \\
$f_{20} - f_{12} = 44\,944.338(72)$~kHz \\
$f_{21} - f_{12} = 44\,996.486(61)$~kHz \\
$\delta(E_4 (v=0,L=0)) = 0.000(860)$~kHz \\
$\delta(E_5 (v=0,L=0)) = 0.000(84)$~kHz \\
$\delta(E_2 (v=0,L=1)) = 0.0000(17)$~kHz \\
$\delta(E_8 (v=0,L=1)) = 0.00000(16)$~kHz \\
$\delta(E_9 (v=0,L=1)) = 0.00000(55)$~kHz \\
\hline
\end{tabular}
\caption{Input data for the least-squares adjustment of the $E_1$, $E_6$, and $E_7$ coefficients, where the line 12 is chosen as reference line.\label{table-LSA}}
\end{table}

We repeated this procedure for all possible choices of a reference line, and found that the adjusted values slightly depend on the chosen line. In our final results (reported in Table~\ref{comparison-exp}), we increased the error bars to make the values compatible with all possible choices. We also checked that the dependence of our results on the hypotheses made on correlations between theoretical errors (see Appendix~\ref{correl}) is negligibly small.

It is worth stressing that the adjustments give satisfactory results (with residuals below $1\sigma$), indicating that experimental data are well explained by the effective spin Hamiltonian introduced in~\cite{Bakalov06} and in good agreement with the theoretical values of the six other coefficients. The only exception is the hyperfine interval involving line 16 (such as $f_{16} - f_{12}$ in Table~\ref{table-LSA}), which has $2-3\sigma$ residuals in all cases. This might be an indication of some experimental problem for this specific line.

\section{Numerical results} \label{SM}

We give here the numerical values of all contributions of orders $m\alpha^6$ and $m\alpha^7 \ln(\alpha)$ to the spin-orbit (Table~\ref{E1table-hdplus-extended}) and spin-spin tensor (Tables~\ref{E6results-extended} and \ref{E7results-extended}) coefficients for a range of rovibrational states ($v=0-9,L=1-4$) of HD$^+$.

\begin{table} [h!]
\small
\begin{tabular}{|@{\hspace{1mm}}c@{\hspace{1mm}}|@{\hspace{1mm}}c@{\hspace{1mm}}|@{\hspace{1mm}}c@{\hspace{1mm}}|@{\hspace{1mm}}c@{\hspace{1mm}}|@{\hspace{1mm}}c@{\hspace{1mm}}
|@{\hspace{1mm}}c@{\hspace{1mm}}|@{\hspace{1mm}}c@{\hspace{1mm}}|@{\hspace{1mm}}c@{\hspace{1mm}}|@{\hspace{1mm}}c@{\hspace{1mm}}
|@{\hspace{1mm}}c@{\hspace{1mm}}|@{\hspace{1mm}}c@{\hspace{1mm}}|@{\hspace{1mm}}c@{\hspace{1mm}}|@{\hspace{1mm}}c@{\hspace{1mm}}|@{\hspace{1mm}}c@{\hspace{1mm}}|}
\hline
$(L,v)$ &$\displaystyle E_1^{(BP)}$& $\displaystyle\mathcal{U}_{Y_1}$ & $\displaystyle\mathcal{U}_{W}$ & $\displaystyle\mathcal{U}_{CM}$ & $\displaystyle\Delta E_{so\hbox{-}H_B}$ &$\displaystyle\Delta E_{so\hbox{-}so}^{(1)}$& $\displaystyle\Delta E_{so\hbox{-}ret}$ & $\displaystyle
\Delta E_1^{(6)}$ & $\displaystyle\mathcal{U}_{Y_2}$ & $\displaystyle\mathcal{U}_{q^2}$ & $\displaystyle \Delta E_{so\hbox{-}H_{(5)}}^{\ln}$ & $\displaystyle
\Delta E_1^{(7\ln)}$ & $\displaystyle E_1$ (this work) \\[2mm]
\hline
(1,0) & 31984.645 & 1.170 & -2.736 & 0.021 & 2.088 & 0.312 & 0.257 & 1.112 & -0.026 & 0.045 & -0.367 & -0.348 & 31985.41(12) \\
(1,1) & 30280.027 & 1.110 & -2.631 & 0.027 & 1.986 & 0.299 & 0.243 & 1.035 & -0.025 & 0.044 & -0.345 & -0.326 & 30280.74(11) \\
(1,2) & 28644.571 & 1.051 & -2.525 & 0.032 & 1.886 & 0.286 & 0.230 & 0.960 & -0.024 & 0.042 & -0.325 & -0.306 & 28645.23(10) \\
(1,3) & 27070.453 & 0.993 & -2.419 & 0.036 & 1.789 & 0.269 & 0.217 & 0.886 & -0.022 & 0.040 & -0.305 & -0.288 & 27071.05(10) \\
(1,4) & 25550.256 & 0.941 & -2.313 & 0.039 & 1.689 & 0.257 & 0.205 & 0.819 & -0.021 & 0.038 & -0.286 & -0.269 & 25550.81(9)  \\
(1,5) & 24076.867 & 0.893 & -2.205 & 0.042 & 1.592 & 0.233 & 0.193 & 0.747 & -0.020 & 0.036 & -0.268 & -0.252 & 24077.36(8)  \\
(1,6) & 22643.449 & 0.834 & -2.097 & 0.044 & 1.498 & 0.219 & 0.181 & 0.679 & -0.019 & 0.035 & -0.251 & -0.235 & 22643.89(8)  \\
(1,7) & 21243.261 & 0.781 & -1.988 & 0.045 & 1.403 & 0.207 & 0.170 & 0.617 & -0.018 & 0.033 & -0.235 & -0.220 & 21243.66(7)  \\
(1,8) & 19869.606 & 0.727 & -1.877 & 0.045 & 1.311 & 0.193 & 0.159 & 0.558 & -0.017 & 0.031 & -0.219 & -0.204 & 19869.96(7)  \\
(1,9) & 18515.725 & 0.669 & -1.764 & 0.045 & 1.217 & 0.179 & 0.147 & 0.493 & -0.016 & 0.029 & -0.203 & -0.190 & 18516.03(6)  \\[1mm]
(2,0) & 31840.910 & 1.164 & -2.719 & 0.020 & 2.069 & 0.311 & 0.256 & 1.101 & -0.026 & 0.045 & -0.364 & -0.345 & 31841.67(12) \\
(2,1) & 30142.781 & 1.105 & -2.614 & 0.026 & 1.969 & 0.296 & 0.243 & 1.024 & -0.025 & 0.043 & -0.343 & -0.324 & 30143.48(11) \\
(2,2) & 28513.394 & 1.047 & -2.509 & 0.031 & 1.868 & 0.281 & 0.229 & 0.948 & -0.023 & 0.042 & -0.322 & -0.304 & 28514.04(10) \\
(2,3) & 26944.952 & 0.992 & -2.404 & 0.035 & 1.771 & 0.266 & 0.216 & 0.877 & -0.022 & 0.040 & -0.303 & -0.285 & 26945.54(10) \\
(2,4) & 25430.065 & 0.937 & -2.298 & 0.039 & 1.673 & 0.252 & 0.204 & 0.807 & -0.021 & 0.038 & -0.284 & -0.267 & 25430.61(9)  \\
(2,5) & 23961.639 & 0.889 & -2.191 & 0.041 & 1.558 & 0.239 & 0.192 & 0.729 & -0.020 & 0.036 & -0.266 & -0.250 & 23962.12(8)  \\
(2,6) & 22532.852 & 0.836 & -2.083 & 0.043 & 1.463 & 0.224 & 0.180 & 0.664 & -0.019 & 0.034 & -0.249 & -0.233 & 22533.28(8)  \\
(2,7) & 21136.966 & 0.785 & -1.974 & 0.044 & 1.369 & 0.211 & 0.169 & 0.604 & -0.018 & 0.032 & -0.232 & -0.217 & 21137.35(7)  \\
(2,8) & 19767.295 & 0.733 & -1.863 & 0.044 & 1.270 & 0.197 & 0.158 & 0.539 & -0.017 & 0.031 & -0.216 & -0.202 & 19767.63(7)  \\
(2,9) & 18417.096 & 0.669 & -1.751 & 0.044 & 1.184 & 0.185 & 0.146 & 0.477 & -0.015 & 0.029 & -0.201 & -0.187 & 18417.39(6)  \\[1mm]
(3,0) & 31627.352 & 1.156 & -2.694 & 0.019 & 2.042 & 0.308 & 0.255 & 1.086 & -0.026 & 0.045 & -0.360 & -0.341 & 31628.10(11) \\
(3,1) & 29938.871 & 1.097 & -2.590 & 0.025 & 1.942 & 0.293 & 0.241 & 1.008 & -0.025 & 0.043 & -0.339 & -0.320 & 29939.56(11) \\
(3,2) & 28318.505 & 1.040 & -2.486 & 0.030 & 1.843 & 0.279 & 0.228 & 0.934 & -0.023 & 0.041 & -0.319 & -0.301 & 28319.14(10) \\
(3,3) & 26758.499 & 0.985 & -2.381 & 0.034 & 1.745 & 0.264 & 0.215 & 0.863 & -0.022 & 0.039 & -0.299 & -0.282 & 26759.08(9)  \\
(3,4) & 25251.499 & 0.930 & -2.275 & 0.037 & 1.649 & 0.250 & 0.203 & 0.795 & -0.021 & 0.038 & -0.281 & -0.264 & 25252.03(9)  \\
(3,5) & 23790.452 & 0.880 & -2.169 & 0.040 & 1.549 & 0.237 & 0.191 & 0.728 & -0.020 & 0.036 & -0.263 & -0.247 & 23790.93(8)  \\
(3,6) & 22368.539 & 0.830 & -2.062 & 0.042 & 1.455 & 0.223 & 0.179 & 0.666 & -0.019 & 0.034 & -0.246 & -0.231 & 22368.98(8)  \\
(3,7) & 20979.051 & 0.780 & -1.953 & 0.043 & 1.356 & 0.209 & 0.168 & 0.602 & -0.018 & 0.032 & -0.229 & -0.215 & 20979.44(7)  \\
(3,8) & 19615.310 & 0.731 & -1.844 & 0.043 & 1.256 & 0.195 & 0.157 & 0.540 & -0.016 & 0.030 & -0.213 & -0.199 & 19615.65(7)  \\
(3,9) & 18270.560 & 0.680 & -1.732 & 0.043 & 1.158 & 0.182 & 0.146 & 0.478 & -0.015 & 0.028 & -0.198 & -0.185 & 18270.85(6)  \\[1mm]
(4,0) & 31346.356 & 1.146 & -2.662 & 0.018 & 2.007 & 0.305 & 0.253 & 1.067 & -0.026 & 0.044 & -0.355 & -0.336 & 31347.09(11) \\
(4,1) & 29670.579 & 1.087 & -2.558 & 0.023 & 1.908 & 0.290 & 0.239 & 0.990 & -0.024 & 0.042 & -0.334 & -0.316 & 29671.25(11) \\
(4,2) & 28062.088 & 1.031 & -2.455 & 0.028 & 1.810 & 0.276 & 0.226 & 0.916 & -0.023 & 0.041 & -0.314 & -0.296 & 28062.71(10) \\
(4,3) & 26513.184 & 0.975 & -2.351 & 0.032 & 1.715 & 0.261 & 0.213 & 0.846 & -0.022 & 0.039 & -0.295 & -0.278 & 26513.96(9)  \\
(4,4) & 25016.561 & 0.923 & -2.246 & 0.036 & 1.618 & 0.248 & 0.201 & 0.779 & -0.021 & 0.037 & -0.276 & -0.260 & 25017.25(9)  \\
(4,5) & 23565.204 & 0.873 & -2.141 & 0.038 & 1.518 & 0.234 & 0.189 & 0.712 & -0.020 & 0.035 & -0.259 & -0.243 & 23565.83(8)  \\
(4,6) & 22152.327 & 0.822 & -2.034 & 0.040 & 1.427 & 0.220 & 0.178 & 0.653 & -0.018 & 0.034 & -0.242 & -0.227 & 22152.87(8)  \\
(4,7) & 20771.242 & 0.772 & -1.927 & 0.041 & 1.328 & 0.207 & 0.166 & 0.588 & -0.017 & 0.032 & -0.226 & -0.211 & 20771.71(7)  \\
(4,8) & 19415.281 & 0.722 & -1.818 & 0.041 & 1.232 & 0.193 & 0.155 & 0.527 & -0.016 & 0.030 & -0.210 & -0.196 & 19415.67(7)  \\
(4,9) & 18077.684 & 0.674 & -1.707 & 0.041 & 1.138 & 0.180 & 0.144 & 0.471 & -0.015 & 0.028 & -0.194 & -0.181 & 18078.00(6)  \\
\hline
\end{tabular}
\caption{Numerical results for the spin-orbit coefficient $E_1$ in HD$^+$ for the  range of rovibrational states $(L=1-4)$ and $(v=0-9)$ (in kHz). All definitions are identical to those given in Table~\ref{E1table-hdplus}. \label{E1table-hdplus-extended}}
\end{table}

\newpage
\begin{table} [h!]
\small
\begin{tabular}{|@{\hspace{1mm}}c@{\hspace{1mm}}|@{\hspace{1mm}}c@{\hspace{1mm}}|@{\hspace{1mm}}c@{\hspace{1mm}}|@{\hspace{1mm}}c@{\hspace{1mm}}|@{\hspace{1mm}}c@{\hspace{1mm}}|@{\hspace{1mm}}c@{\hspace{1mm}}
|@{\hspace{1mm}}c@{\hspace{1mm}}|@{\hspace{1mm}}c@{\hspace{1mm}}|}
\hline
$(L,v)$ & $\displaystyle E_6^{(BP)}$ & $\displaystyle\mathcal{U}_{2d}^{(2)}$&$\displaystyle\mathcal{U}_{5b}^{(2)}$ & $\displaystyle\Delta E_{ss}^{(2)}$ & $\displaystyle\Delta E_6^{(6)}$ &  $\displaystyle\Delta E_6^{(7)} $ & $\displaystyle E_6$ (this work)  \\[2mm]
\hline
(1,0) & 8611.112 & -0.806 & 0.093 & 0.955 & 0.242 & -0.055 & 8611.299(18) \\
(1,1) & 8136.686 & -0.770 & 0.090 & 0.905 & 0.225 & -0.052 & 8136.859(17) \\
(1,2) & 7682.610 & -0.735 & 0.087 & 0.857 & 0.210 & -0.049 & 7682.771(16) \\
(1,3) & 7246.701 & -0.700 & 0.084 & 0.807 & 0.191 & -0.047 & 7246.846(16) \\
(1,4) & 6826.898 & -0.666 & 0.082 & 0.762 & 0.178 & -0.044 & 6827.031(15) \\
(1,5) & 6421.232 & -0.632 & 0.079 & 0.726 & 0.172 & -0.041 & 6421.364(14) \\
(1,6) & 6027.809 & -0.599 & 0.076 & 0.678 & 0.154 & -0.039 & 6027.925(13) \\
(1,7) & 5644.779 & -0.565 & 0.071 & 0.633 & 0.140 & -0.037 & 5644.883(12) \\
(1,8) & 5270.312 & -0.531 & 0.068 & 0.590 & 0.127 & -0.034 & 5270.405(11) \\
(1,9) & 4902.573 & -0.498 & 0.065 & 0.541 & 0.108 & -0.032 & 4902.649(11) \\[1mm]
(2,0) & 2043.1452 & -0.1908 & 0.0219 & 0.2259 & 0.0569 & -0.0130 & 2043.1891(43) \\
(2,1) & 1930.5230 & -0.1823 & 0.0213 & 0.2144 & 0.0534 & -0.0124 & 1930.5640(41) \\
(2,2) & 1822.7196 & -0.1740 & 0.0207 & 0.2032 & 0.0499 & -0.0117 & 1822.7578(39) \\
(2,3) & 1719.2168 & -0.1658 & 0.0200 & 0.1911 & 0.0454 & -0.0110 & 1719.2512(37) \\
(2,4) & 1619.5252 & -0.1576 & 0.0193 & 0.1806 & 0.0422 & -0.0105 & 1619.5570(35) \\
(2,5) & 1523.1776 & -0.1496 & 0.0185 & 0.1712 & 0.0401 & -0.0098 & 1523.2080(33) \\
(2,6) & 1429.7235 & -0.1416 & 0.0178 & 0.1577 & 0.0339 & -0.0092 & 1429.7482(31) \\
(2,7) & 1338.7229 & -0.1337 & 0.0170 & 0.1471 & 0.0304 & -0.0086 & 1338.7446(39) \\
(2,8) & 1249.7406 & -0.1257 & 0.0162 & 0.1353 & 0.0257 & -0.0081 & 1249.7582(27) \\
(2,9) & 1162.3394 & -0.1177 & 0.0153 & 0.1257 & 0.0233 & -0.0075 & 1162.3552(25) \\[1mm]
(3,0) &  948.5222 & -0.0883 & 0.0101 & 0.1042 & 0.0260 & -0.0060 &  948.5421(20) \\
(3,1) &  896.1987 & -0.0843 & 0.0098 & 0.0992 & 0.0247 & -0.0057 &  896.2176(19) \\
(3,2) &  846.1054 & -0.0805 & 0.0095 & 0.0936 & 0.0227 & -0.0054 &  846.1227(18) \\
(3,3) &  798.0019 & -0.0767 & 0.0092 & 0.0880 & 0.0206 & -0.0051 &  798.0174(17) \\
(3,4) &  751.6607 & -0.0729 & 0.0089 & 0.0833 & 0.0193 & -0.0048 &  751.6752(16) \\
(3,5) &  706.8644 & -0.0692 & 0.0086 & 0.0793 & 0.0187 & -0.0045 &  706.8786(15) \\
(3,6) &  663.4035 & -0.0655 & 0.0082 & 0.0737 & 0.0165 & -0.0042 &  663.4158(14) \\
(3,7) &  621.0731 & -0.0618 & 0.0079 & 0.0690 & 0.0151 & -0.0040 &  621.0842(13) \\
(3,8) &  579.6702 & -0.0581 & 0.0075 & 0.0645 & 0.0138 & -0.0037 &  579.6803(12) \\
(3,9) &  538.9906 & -0.0544 & 0.0070 & 0.0593 & 0.0119 & -0.0035 &  538.9991(12) \\[1mm]
(4,0) &  550.5184 & -0.0510 & 0.0058 & 0.0600 & 0.0148 & -0.0035 &  550.5298(12) \\
(4,1) &  520.1196 & -0.0487 & 0.0056 & 0.0570 & 0.0139 & -0.0033 &  520.1302(11) \\
(4,2) &  491.0099 & -0.0465 & 0.0055 & 0.0536 & 0.0126 & -0.0031 &  491.0194(10) \\
(4,3) &  463.0496 & -0.0443 & 0.0053 & 0.0508 & 0.0118 & -0.0029 &  463.0585(10) \\
(4,4) &  436.1067 & -0.0421 & 0.0052 & 0.0480 & 0.0110 & -0.0028 &  436.1150(9)  \\
(4,5) &  410.0548 & -0.0399 & 0.0049 & 0.0446 & 0.0096 & -0.0026 &  410.0618(9)  \\
(4,6) &  384.7717 & -0.0378 & 0.0047 & 0.0424 & 0.0094 & -0.0024 &  384.7786(8) \\
(4,7) &  360.1380 & -0.0356 & 0.0045 & 0.0398 & 0.0086 & -0.0023 &  360.1443(8) \\
(4,8) &  336.0351 & -0.0335 & 0.0043 & 0.0370 & 0.0078 & -0.0022 &  336.0407(7) \\
(4,9) &  312.3437 & -0.0313 & 0.0040 & 0.0335 & 0.0063 & -0.0020 &  312.3480(7) \\
\hline
\end{tabular}
\caption{Numerical results for the electron-proton spin-spin tensor coefficient $E_6$ in HD$^+$ for the range of rovibrational states $(L=1-4)$ and $(v=0-9)$ (in kHz). All definitions are identical to those given in Table~\ref{E6results}. \label{E6results-extended}}
\end{table}

\newpage
\begin{table} [h!]
\small
\begin{tabular}{|@{\hspace{1mm}}c@{\hspace{1mm}}|@{\hspace{1mm}}c@{\hspace{1mm}}|@{\hspace{1mm}}c@{\hspace{1mm}}|@{\hspace{1mm}}c@{\hspace{1mm}}
|@{\hspace{1mm}}c@{\hspace{1mm}}|@{\hspace{1mm}}c@{\hspace{1mm}}|@{\hspace{1mm}}c@{\hspace{1mm}}|@{\hspace{1mm}}c@{\hspace{1mm}}|}
\hline
$(L,v)$ & $\displaystyle E_7^{(BP)}$ & $\displaystyle\mathcal{U}_{2d}^{(2)}$&$\displaystyle\mathcal{U}_{5b}^{(2)}$ & $\displaystyle\Delta E_{ss}^{(2)}$ & $\displaystyle\Delta E_7^{(6)}$ &  $\displaystyle\Delta E_7^{(7)} $ &
$\displaystyle E_7$ (this work) \\[2mm]
\hline
(1,0) & 1321.7673 & -0.1237 & 0.0142 & 0.1467 & 0.0371 & -0.0085 & 1321.7960(28) \\
(1,1) & 1248.9359 & -0.1182 & 0.0138 & 0.1390 & 0.0346 & -0.0080 & 1248.9624(27) \\
(1,2) & 1179.2284 & -0.1128 & 0.0135 & 0.1316 & 0.0322 & -0.0076 & 1179.2530(25) \\
(1,3) & 1112.3096 & -0.1075 & 0.0130 & 0.1239 & 0.0293 & -0.0072 & 1112.3318(24) \\
(1,4) & 1047.8631 & -0.1023 & 0.0125 & 0.1170 & 0.0273 & -0.0068 & 1047.8836(23) \\
(1,5) &  985.5868 & -0.0971 & 0.0121 & 0.1115 & 0.0264 & -0.0063 &  985.6069(21) \\
(1,6) &  925.1895 & -0.0919 & 0.0115 & 0.1040 & 0.0237 & -0.0060 &  925.2072(20) \\
(1,7) &  866.3873 & -0.0867 & 0.0111 & 0.0972 & 0.0215 & -0.0056 &  866.4032(19) \\
(1,8) &  808.8992 & -0.0816 & 0.0105 & 0.0906 & 0.0196 & -0.0053 &  808.9135(18) \\
(1,9) &  752.4434 & -0.0764 & 0.0099 & 0.0830 & 0.0166 & -0.0049 &  752.4552(16) \\[1mm]
(2,0) &  696.7322 & -0.0293 & 0.0033 & 0.0347 & 0.0087 & -0.0020 &  313.6203(7) \\
(2,1) &  296.3244 & -0.0280 & 0.0032 & 0.0329 & 0.0082 & -0.0019 &  296.3307(6) \\
(2,2) &  279.7750 & -0.0267 & 0.0031 & 0.0312 & 0.0077 & -0.0018 &  279.7808(6) \\
(2,3) &  263.8857 & -0.0254 & 0.0030 & 0.0293 & 0.0070 & -0.0017 &  263.8910(6) \\
(2,4) &  248.5814 & -0.0242 & 0.0029 & 0.0277 & 0.0065 & -0.0016 &  248.5863(5) \\
(2,5) &  233.7905 & -0.0230 & 0.0028 & 0.0263 & 0.0062 & -0.0015 &  233.7951(5) \\
(2,6) &  219.4436 & -0.0217 & 0.0027 & 0.0242 & 0.0052 & -0.0014 &  219.4474(5) \\
(2,7) &  205.4734 & -0.0205 & 0.0026 & 0.0226 & 0.0047 & -0.0013 &  205.4767(4) \\
(2,8) &  191.8128 & -0.0193 & 0.0024 & 0.0208 & 0.0040 & -0.0012 &  191.8155(4) \\
(2,9) &  178.3949 & -0.0181 & 0.0023 & 0.0193 & 0.0036 & -0.0011 &  178.3973(4) \\[1mm]
(3,0) &  145.5939 & -0.0136 & 0.0016 & 0.0160 & 0.0040 & -0.0009 &  145.5969(3) \\
(3,1) &  137.5614 & -0.0129 & 0.0016 & 0.0152 & 0.0038 & -0.0009 &  137.5643(3) \\
(3,2) &  129.8713 & -0.0124 & 0.0014 & 0.0144 & 0.0035 & -0.0008 &  129.8740(3) \\
(3,3) &  122.4867 & -0.0118 & 0.0014 & 0.0135 & 0.0032 & -0.0008 &  122.4891(3) \\
(3,4) &  115.3726 & -0.0112 & 0.0013 & 0.0128 & 0.0030 & -0.0007 &  115.3748(2) \\
(3,5) &  108.4956 & -0.0106 & 0.0013 & 0.0122 & 0.0029 & -0.0007 &  108.4978(2) \\
(3,6) &  101.8236 & -0.0100 & 0.0013 & 0.0113 & 0.0025 & -0.0006 &  101.8255(2) \\
(3,7) &   95.3251 & -0.0095 & 0.0012 & 0.0106 & 0.0023 & -0.0006 &   95.3268(2) \\
(3,8) &   88.9689 & -0.0089 & 0.0012 & 0.0099 & 0.0021 & -0.0006 &   88.9705(2) \\
(3,9) &   82.7237 & -0.0083 & 0.0010 & 0.0091 & 0.0018 & -0.0005 &   82.7250(2) \\[1mm]
(4,0) &   84.5020 & -0.0078 & 0.0009 & 0.0092 & 0.0023 & -0.0005 &   84.5038(2) \\
(4,1) &   79.8354 & -0.0075 & 0.0009 & 0.0088 & 0.0021 & -0.0005 &   79.8370(2) \\
(4,2) &   75.3666 & -0.0071 & 0.0008 & 0.0082 & 0.0019 & -0.0005 &   75.3680(2) \\
(4,3) &   71.0742 & -0.0068 & 0.0008 & 0.0078 & 0.0018 & -0.0005 &   71.0756(2) \\
(4,4) &   66.9381 & -0.0065 & 0.0007 & 0.0074 & 0.0017 & -0.0004 &   66.9393(1) \\
(4,5) &   62.9387 & -0.0061 & 0.0007 & 0.0068 & 0.0015 & -0.0004 &   62.9397(1) \\
(4,6) &   59.0573 & -0.0058 & 0.0007 & 0.0065 & 0.0014 & -0.0004 &   59.0583(1) \\
(4,7) &   55.2755 & -0.0055 & 0.0007 & 0.0061 & 0.0013 & -0.0004 &   55.2765(1) \\
(4,8) &   51.5753 & -0.0051 & 0.0006 & 0.0057 & 0.0012 & -0.0003 &   51.5761(1) \\
(4,9) &   47.9381 & -0.0048 & 0.0006 & 0.0051 & 0.0010 & -0.0003 &   47.9388(1) \\
\hline
\end{tabular}
\caption{Numerical results for the electron-deuteron spin-spin tensor coefficient $E_7$ in HD$^+$ for the range of rovibrational states $(L=1-4)$ and $(v=0-9)$ (in kHz). All definitions are identical to those given in Table~\ref{E6results}. \label{E7results-extended}}
\end{table}

\end{document}